# INTERACTIVE DECISION SUPPORT SYSTEM FOR LUNG CANCER SEGMENTATION

## V. SYDORSKYI

**Abstract.** This paper studies Clinical Intelligent Decision Support Systems (CIDSSs) for lung cancer segmentation, which are based on deep neural nets. A new interactive CIDSS is proposed and compared with previous approaches. Additionally, the purpose uncertainty problem in building interactive systems is discussed, and criteria for measuring both quality and amount of user feedback are proposed. In order to automate system evaluation, a new algorithm was used to simulate expert feedback. The proposed interactive CIDSS outperforms previous approaches (both interactive and noninteractive) on the task of lung lesion segmentation. This approach looks promising both in terms of quality and expert user experience. At the same time, this paper discusses a bunch of possible modifications that can be done to improve both evaluation criteria and proposed CIDSS in future works.

**Keywords:** clinical decision support systems, deep learning, open system, interactive segmentation.

## INTRODUCTION

Decision support systems (DSS) stand out as instrumental across various domains of human activities [1–3]. These systems combine the advantages of big data, statistical models, big informational systems, classical machine learning, and deep learning technologies. Of particular academic and practical interest are the emerging Intelligent DSS. Characterized by their reliance on neural networks, these systems promise enhanced analytical depth and precision, setting the stage for transformative applications across diverse sectors.

A significant portion of recent research [4] has been dedicated to the application of machine learning and deep learning in medical imaging. These studies primarily focus on architecting neural networks and formulating optimization policies to improve diagnostic accuracy. However, it is important to consider deep learning models as part of DSS systems and sub-systems in medical applications. Being a part of IDSS, they can benefit from advances in computer science, system analysis, and decision-making in scientific and practical spheres. IDDS, based on neural nets, can provide accurate medical insights, a user-friendly interface, and an interactive and adaptive mechanism for decision-making.

To address the problem of building IDSS based on deep learning approaches, using expert feedback to improve the initial results of the proposed system, the task of lung cancer segmentation is considered. The main contributions of this paper are:

- Integration of a segmentation neural net into an interactive intelligent decision support system.

• Adaptation of the previous interactive segmentation approach [5] to the task of lung cancer segmentation.

• Improvement of the previous approach by the usage of two types of segmentation neural nets.

• Formulate a purpose uncertainty problem [6] in the construction of such a DSS and propose a set of criteria for DSS assessment, including a special algorithm for expert feedback simulation.

Also, it is important to mention that the proposed system can be adapted for other types of cancer diagnosis. Such adaptions will be considered in future works.

**RELATED WORK**

The application of clinical decision support systems (CDSS) was proposed in recent studies in order to help with the diagnosis and treatment of oncology diseases [7], modifications in order to adapt to cancer treatment in developing countries [8], and specialized DSS, for example for brain tumors treatment [9]. All these works are focused on building complete decision support systems. Still, at the same time, they focus less on particular intelligent subsystems, which hugely benefit the precision of the overall system.

The latest research in computer science proposes a wide variety of deep learning algorithms [10], which can be utilized in order to improve the quality of CDSS. Starting from classical U-Net [11], FPN [12], and DeepLab [13] architectures, which made a breakthrough in semantic image segmentation in the sphere of medical imaging, ending with the latest Unet++ [14], UNet 3+ [15] and UNETR [16] architectures, which rely on complex skip and residual connections and on processing the whole image volume altogether (3D Nets). Additionally, different Hybrid and Neuro-Fuzzy Networks were applied to the task of lung and brain tumor segmentation and detection in combination with classical 2D, MIL, and 3D approaches [17].

At the same time, interactive segmentation methods propose guiding deep learning algorithms with user feedback. Segment Anything Model [18] proved that neural nets can generalize to the segmentation of many different objects in completely different scenarios. However, there is doubt that it can segment precise medical images, especially in the presence of very tiny and hard distinguishable objects, like lung cancer lesions. Other more specialized approaches were proposed [19; 5] and showed success in medical image segmentation.

**INTERACTIVE INTELLIGENT DECISION SUPPORT SYSTEM FOR LUNG CANCER SEGMENTATION**

This work considers modern deep learning algorithms to build an interactive intelligent decision support system for lung cancer segmentation. The proposed approach operates on CT scans from different manufacturers and incorporates user feedback to increase the segmentation quality. Finally, three possible architectures of DSS are compared by several criteria.

First, let's formalize the main problem that the segmentation system should solve. This system receives a pre-processed CT scan — $x$, which is a 3D image, where dimensions refer to the width, height, and depth of the image matrix:

$$x = \{x_{ijk}\}, \quad i = 0,...,height, \quad j = 0,...,width, \quad k = 0,...,n\_slices.$$

In the proposed approach, the Multi-Instance Learning (MIL) [17; 20] approach is used, where each instance refers to a slice of the scan. So, the basic deep learning model — $H$ should approximate mask pixel distribution for each slice based on several scan slices:

$$H(x_K) = \widehat{y_k}, \quad x_K = x_{IJK}, \quad \widehat{y_k} = \widehat{y_{IJk}},$$

$$I = \{0,1,...,height\}, \quad J = \{0,1,...,width\}, \quad K = \{k_1, k_2,...,k_l\},$$

where $K$ refers to some set of indexes that define slices, which are used for the prediction of each slice mask, $k$ refers to some fixed slice index, and $\widehat{y_k}$ refers to an approximated slice mask.

In order to make the system interactive, it is required to introduce additional input to the network, which refers to the expert feedback on initially approximated slice masks. One of the ways to incorporate such feedback is to allow expert to select pixels that, he thinks, are misclassified. An expert can simply click on pixels that he thinks should be assigned to foreground — lesion region or background — non lesion region. Also, an expert can erase the whole slice mask. Formalization of expert's feedback:

• Two sets of clicks are provided by the expert: $c\_p$ — "positive" clicks, which refer to the foreground ($c\_p = \{i, j\}$, $i \in \{0,1,...,height\}$, $j \in \{0,1,...,width\}$) and $c\_n$ — "negative" clicks, which refer to the background ($c\_n = \{i, j\}$, $i \in \{0,1,...,height\}$, $j \in \{0,1,...,width\}$).

• Click mask — $cm$ ($cm = \{cm_{ij}\}$, $i = 0,...,height$, $j = 0,...,width$, $cm_{ij} \in [0,+\infty)$).

• Algorithm for encoding clicks into click mask $C2M$ ($C2M(c\_p) = cm\_p$ OR $C2M(c\_n) = cm\_n$).

• New model — $H\_I$ ($H\_I(cm\_p, cm\_n, \widehat{y_k}, x_K) = \widehat{y\_f_k}$).

It is important to mention that $H\_I$ takes the previous predicted mask $\widehat{y_k}$ on input, which allows to incorporate both information from the previous step and expert feedback. The interactive procedure can be applied several times: redefine $\widehat{y_k} := \widehat{y\_f_k}$ and ask expert for new $c\_p$ and $c\_n$. Such an iterative procedure creates a certain trade-off: more iterations result in better segmentation masks while taking more expert's time. Also, to receive the initially predicted mask, two approaches can be used: predicted masks from $H$ or initialize $\widehat{y_k}$ as zero mask and ask an expert to define foreground areas without the initially predicted mask.

So there are three possible structures of segmentation DSS:

1. Noninteractive (Fig. 1).

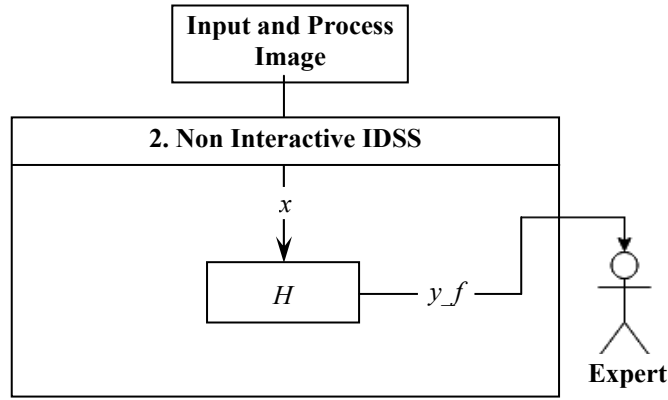

*Fig. 1.* Structure of noninteractive IDSS

2. Interactive, without the initially predicted mask (Fig. 2).

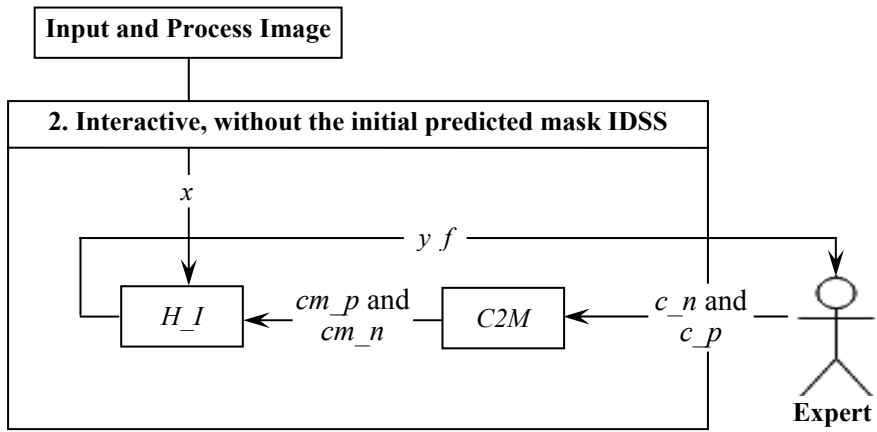

*Fig. 2.* Structure of interactive IDSS, without the initially predicted mask

3. Interactive with initially predicted mask, received from $H$ (Fig. 3).

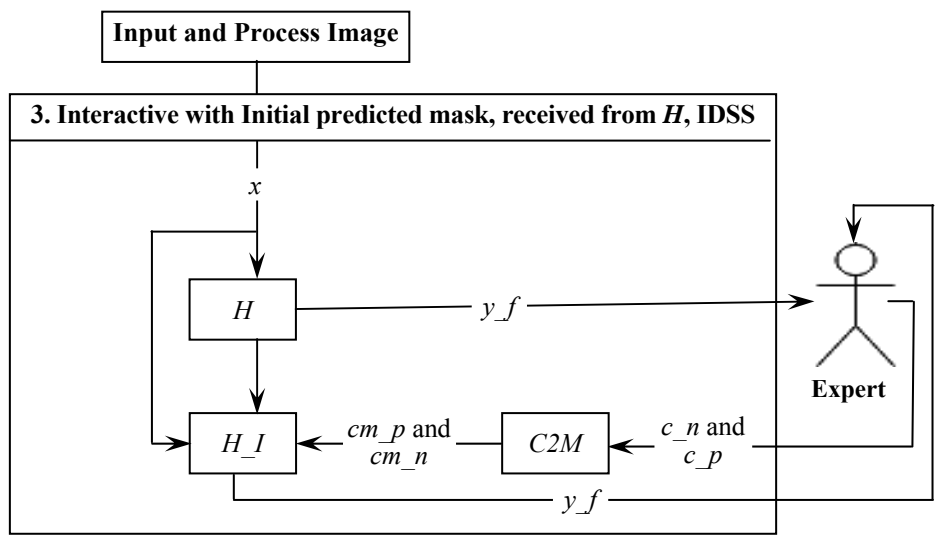

*Fig. 3.* Structure of interactive IDSS with initially predicted mask received from $H$

Although, it is important to mention that proposed segmentation DSSs are open systems because they explicitly have information exchange in the form of user feedback. While utilizing this user feedback, entropy in the system is decreasing. This can be easily tracked by the improvement of quality metrics, which will be discussed in Experiments section.

**Purpose uncertainty problem.** One of the problems in building such an IDSS is purpose uncertainty [6] because there is a trade-off between the amount of user interaction and the quality of the final mask. In this task, different ways of expert interaction require different amounts of effort:

– Making a positive click is the most challenging action because it requires an understanding of precise lesion location, while the initial mask, produced by DSS, can bias the expert.

– Making negative clicks requires less effort because, in most cases, finding a healthy region is an easier task.

– Erasing the whole mask is the most straightforward action because it does not require selecting a particular region.

Considering quality metrics, there is also some uncertainty because the proposed systems will be tested on a series of CT scans, but it is important to track both average quality results and a number of poorly segmented masks. In order to handle this problem, median metric values and the whole metric distribution will be considered in Evaluation section.

**DETAILED NEURAL NET SETUP**

First of all, $K$ set for selecting input instances should be defined. In the proposed approach $K = \{k-2, \ k-1, \ k, \ k+1, \ k+2\}$, $k$ refers to target slice.

So, the neural net will operate on the target slice and two previous and next slices. Initial experimental results proved that using more slices does not improve segmentation quality and only leads to optimization and inference overheads and overfitting. To feed selected slices in regular 2D Convolutions, slices will be stacked on channel dimension so the resulting input will have the following form:

$$x\_k = x\_KIJ, \ I = \{0,1,...,height\}, \ J = \{0,1,...,widt\}.$$

In this paper, classical Encoder-Decoder segmentation neural nets will be used: EfficientNet B3 [21] encoder and Unet++ [14], DeepLabV3+ [22] decoders. Different setups will be evaluated in Experiments section, and the best one will be chosen according to the proposed criteria. After the final fully connected layer, the sigmoid will constraint outputs of each pixel prediction to [0; 1] range, and thresholding by 0.5 will be applied.

To design click-to-mask transformation, "Gaussian" masks will be used, as proposed in [5] — equations:

$$DM_{c_l} = \{d_{ij}\}, \ d_{ij} = d((i,j), c_l);$$

$$CM_{c_l} = \{cm_{ij}\}, \ cm_{ij} = e^{(-d_{ij}^2/(2sigma^2))} \ \text{if} \ d_{ij} < clip\_radius \ \text{else} \ 0;$$

$$CM = \sum_{c_l} CM_{cl},$$

where $DM_{c_l}$ is a distance transform for each click $c_l$; $d$ — euclidean distance; $CM_{c_l}$ — click mask for each click $c_l$; *sigma* — "Gaussian standard deviation"; *clip_radius* parameter, which zeros out mask pixels, which are too far from the center; $CM$ final click mask for several clicks. In the proposed approach, separate masks for positive and negative clicks are used. Finally, to feed clicks masks and previously predicted masks to an interactive neural net $H\_I$, all inputs are stacked channel-wise, so $H\_I$ has three more additional input channels compared to $H$.

**DETAILED OPTIMIZATION SETUP**

For training both $H$ and $H\_I$ Adam optimizer [23] and reducing the learning rate with the OneCycleCosine learning rate scheduler, starting with 1e-3 and finishing with 1e-6, are used. Neural nets are trained for 191000 steps. Batch size is set to 32, and a distributed data parallel [24] mechanism on two A100 GPUs is used.

It is important to consider the data distribution of masked regions, which is typical for most medical segmentation tasks. In most cases, regions with lesions take only a very small partition of the whole image area, so there are many more negative pixels than positive ones. This work proposes to sample slices with and without lesions with equal probability to reduce class imbalance for the optimization procedure.

In order to optimize both neural nets, the sum of BCE and Jaccard [25] losses is used.

Also, a transfer learning mechanism is used, and the encoder is initialized with weights received from a noisy student learning procedure [26].

For training noninteractive net ($H$) horizontal and vertical flip augmentations are used, and on the inference stage, the same test time augmentation [27] is applied to improve prediction quality. For training interactive net ($H\_I$) any augmentations are not used.

Another important part is the click-sampling strategy. This work mostly follows strategies proposed in [5] but with several modifications specific to the medical imaging domain. Initially, there are not previously predicted masks, so first clicks should be initialized by some "cold start" strategy. For positive clicks, the center masses of each lesion are selected; if there are several lesions on one slice, the lesion with the biggest area is picked. For negative clicks, random points within $d_0$ distance from the lesion are selected. Other negative click sampling strategies from [5] were not considered because of task specifications:

• Sampling from another object is not possible in this task because there is only one type of object to segment.

• Sampling from the target object border is also inefficient because the lesion area is pretty tiny, and its border may contain a big part of this area.

After sampling clicks, a click mask for the current interaction can be created and saved to the click cache for the next interactions. Predicted masks for click sampling are used only after the first epoch. Having previously predicted masks and previous clicks, it is possible to use other sampling strategies. For selecting a positive click, the algorithm uses false negative pixels connected regions and selects the biggest area region and sample click, which is located on the largest distance both from the region border and previous clicks:

$$c_l = \arg\max_p(\min(\min_{p_t} d(p, p_r), \min_{p_c} d(p, p_c))), \qquad (1)$$

where $c_l$ refers to click to select; $p$ — all possible points; $p_t$ — points of the region; $p_c$ — previous clicks from the cache. The algorithm follows the same logic for negative clicks, but here, false positive masks are used as mislabeled masks. To introduce diversity and ensure that the neural net can work only with a few clicks, the algorithm randomly resets the click cache for each image with a probability of 0.3. Also, the maximum number of positive and negative clicks is constrained to 12 separately. It is done to reduce RAM overhead and ensure that the neural netdoes not abuse the usage of lots of clicks.

**EXPERIMENTAL RESULTS**

In this section, the following points are described:

– Lung lesions dataset, which is used for training neural networks and testing proposed DSSs;

– Metric that is used for evaluation and solution for purpose uncertainty issue;

– An algorithm for simulating feedback from an expert;

– Results of the proposed approach.

**Dataset**. To experiment with the proposed IDSS, the Lung Image Database Consortium image collection (LIDC-IDRI) [28] is used, which consists of diagnostic and lung cancer screening thoracic computed tomography (CT) scans with marked-up annotated lesions. In this research, only CT scans and marked-up annotated lesions are used.

The whole dataset contains 1018 scans from 1010 unique patients. So, in most cases, there is one CT scan for one patient. The dataset annotation was obtained from four independent experts. To get the final mask, the consensus of the majority of annotators [29] was used. The examples of original and masked images are presented in Fig. 4.

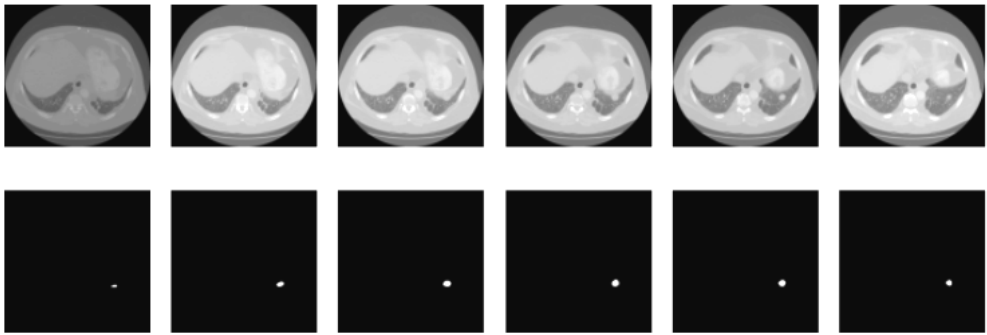

*Fig. 4.* Examples of CT scans with annotation. First row — original scans. Second row — annotation

All CT scans have $512 \times 512$ height and width, but they all differ in the number of slices. Slices distribution can be seen in Fig. 5.

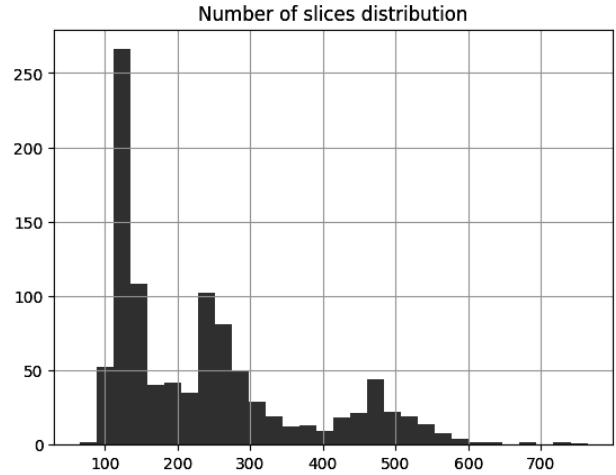

*Fig. 5.* Distribution of the number of slices across the dataset

As mentioned before, the biggest problem in lesion segmentation is a severe class imbalance — most of the pixels are background pixels. Moreover, most of the slices do not contain lesions: the median number of slices in the scan is 203.5, and the median number of slices with lesions is 40. Also, the area of lesions is pretty small; the median of the relative area of the foreground area to the whole scan area across all dataset images is 8.7683e-06. So, such class imbalance should be considered during model optimization and evaluation.

**Evaluation**. Two main criteria to evaluate results will be considered: quality and amount of expert feedback.

For quality evaluation, the Intersection over Union (IoU) [30] metric is used, which is computed for each scan separately and averaged across all scans. If the scan does not contain lesions and the predicted mask is empty — IoU is equal to 1. This work also outlines the median, first ($Q_1$), and third ($Q_3$) quartiles of IoU scores and overall distribution.

For the amount of expert feedback evaluation, three types of feedback, outlined in Purpose uncertainty problem section, are considered. The feedback score equation (2) includes different weights for feedback types. These weights are proposed based on personal perception of the process and can be tuned in future work.

$$Feedback\ Score = Number\ of\ Positive\ Clicks +$$

$$+ Number\ of\ Negative\ clicks\ *0.85 +$$

$$+ Number\ of\ Masks\ erasements\ *0.75\ . \qquad (2)$$

This score should be minimized.

The dataset is split into train and test sets with stratification by relative area of foreground area and grouping by patients (so scans of one patient are only in the test or train set). After the split, there are 203 scans in the test set and 815 scans in the train set, so approximately 20% of the data is used for model assessment.

**Algorithm for simulating expert feedback**. This algorithm is needed to test our second and third types of IDSS. The proposed algorithm behaves as an "ideal" expert. The algorithm makes one optimal click for each slice in each scan on each iteration. Such an approach is impossible in the real scenario, but professional radiologists should behave near the "ideal" behavior. This work proposes possible improvements for future work:

• Introduce the probability of making positive, negative clicks or mask erased.

• Introduce some probability distribution of click coordinate.

• Condition both previous distributions on lesion size, number of ground truth, and predicted lesions in the particular CT scan.

• Reduce the probability of correct feedback with iteration number.

Let's formalize optimal positive and negative clicks. First, false positive pixels connected regions for positive click and false negatives connected regions for negative are computed. If there are no such regions for a particular scan or the whole region consists only of its border (single-pixel line), the algorithm just omits clicking. Otherwise, it selects the biggest mislabeled region using the technique proposed in equation (1).

If no feedback was done for a particular slice, a mask from the previous iteration or an empty mask if we have a "cold start" scenario is used.

**Experiments.** Both different systems and different decoders for neural networks $H$ and $H\_I$ will be compared by evaluation metrics proposed in Evaluation section

Quality metrics are outlined in Table 1.

**T a b l e  1.** Quality metrics

| System and Decoder | *Mean IoU* ↑ | *Median IoU* ↑ | *$Q_1$ IoU* ↑ | *$Q_3$ IoU* ↑ |
|---|---|---|---|---|
| System 1<br>Unet++ Decoder | 0.4083 | 0.4385 | 0.1507 | 0.5993 |
| System 1<br>DeepLabV3+ Decoder | 0.3929 | 0.4104 | 0.1202 | 0.6103 |
| System 2<br>Unet++ Decoder | 0.7031 | 0.6904 | 0.6154 | 0.7729 |
| System 2<br>DeepLabV3+ Decoder | 0.6868 | 0.6679 | 0.5925 | 0.7587 |
| System 3<br>Unet++ Decoder | **0.7182** | **0.7039** | **0.6351** | **0.7945** |
| System 3<br>DeepLabV3+ Decoder | 0.6986 | 0.6833 | 0.6063 | 0.7844 |

DeepLabV3+ and Unet++ decoders are compared because the first one was used in [5]. From the IoU score, it is obvious that Unet++ outperforms the DeepLabV3+ decoder for all systems, which is logical because the model can benefit from lots of skip connections in Unet++ architecture. Bootstrapped cross-entropy [31] loss was also tested in this work, but the results were worse than the proposed loss setup. Comparing different systems, the proposed System 3 outperforms all other systems by all quality metrics and with all proposed decoders. It is essential to pay attention to Mean IoU, and $Q_1$ IoU : we can see that the interactive approach introduces more than 0.3 improvements in mean score and a bit less than 0.5 improvement in $Q_1$ score, which is a huge increase in quality.

Next, we can observe IoU scores distribution for System 1 Unet++ Decoder, System 2 Unet++ Decoder, and System 3 Unet++ Decoder in Fig. 6.

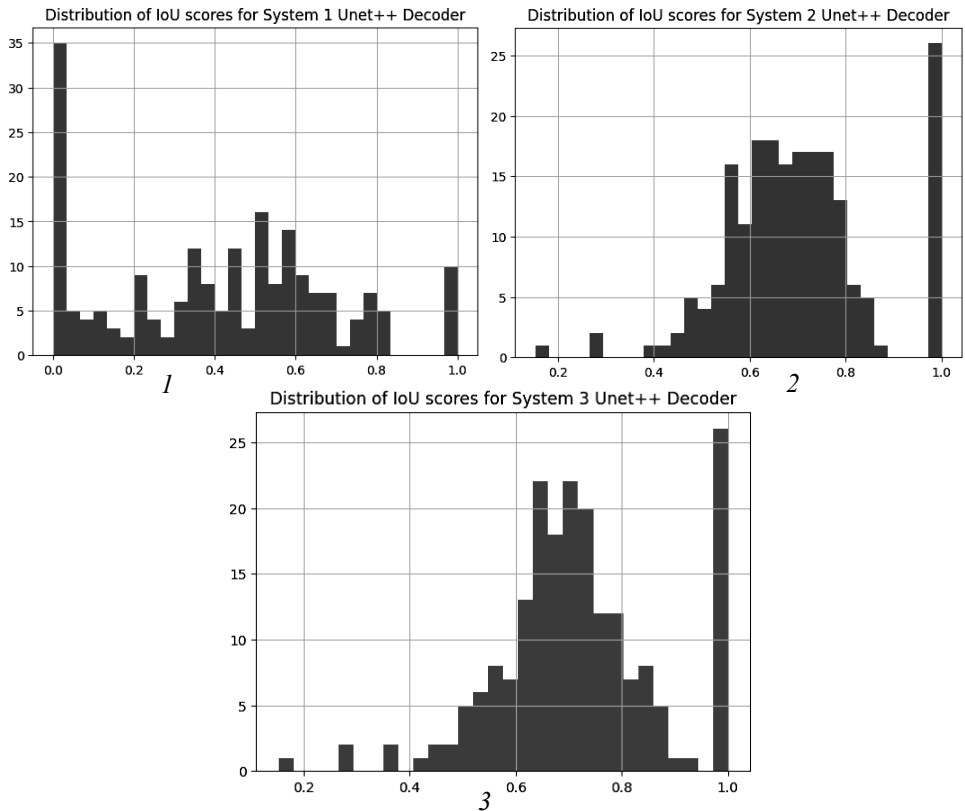

*Fig. 6.* IoU scores distribution for Systems 1, 2, and 3 with Unet++ decoder

Obtained results show fewer scans segmented with very poor scores and many more scans segmented with nearly ideal scores.

Visual results of segmentation can be seen in Figs. 7, 8.

According to visual results, the model correctly reacts to positive and negative clicks. Additionally, the model does not remove the correct segmentation if negative clicks are placed near correctly segmented regions. However, there is still space for improvement. As we can see from Fig. 7, the model is pretty conservative in segmenting additional areas after a positive click for some scans.

Feedback Score (equation (2)) is outlined only for interactive systems (second and third) in Table 2.

**T a b l e  2.** Number of all feedback types and resulting Feedback Score

| System and Decoder | Number of Positive Clicks ↓ | Number of Negative Clicks ↓ | Number of Masks Erasements ↓ | Feedback Score ↓ |
|---|---|---|---|---|
| System 2 Unet++ Decoder | 3024 | **0** | **0** | 3024 |
| System 2 DeepLabV3+ Decoder | 3024 | **0** | **0** | 3024 |
| System 3 Unet++ Decoder | **1637** | 284 | 1384 | **2916.4** |
| System 3 DeepLabV3+ Decoder | 1730 | 248 | 1344 | 2948.8 |

From Table 2: System 2 requires only positive clicks because there is no initial mask approximation, while System 3 requires correction of initial results, so it has all types of feedback. Considering weighting coefficients from equation (2), we see that System 3 again outperforms System 2 with all decoders.

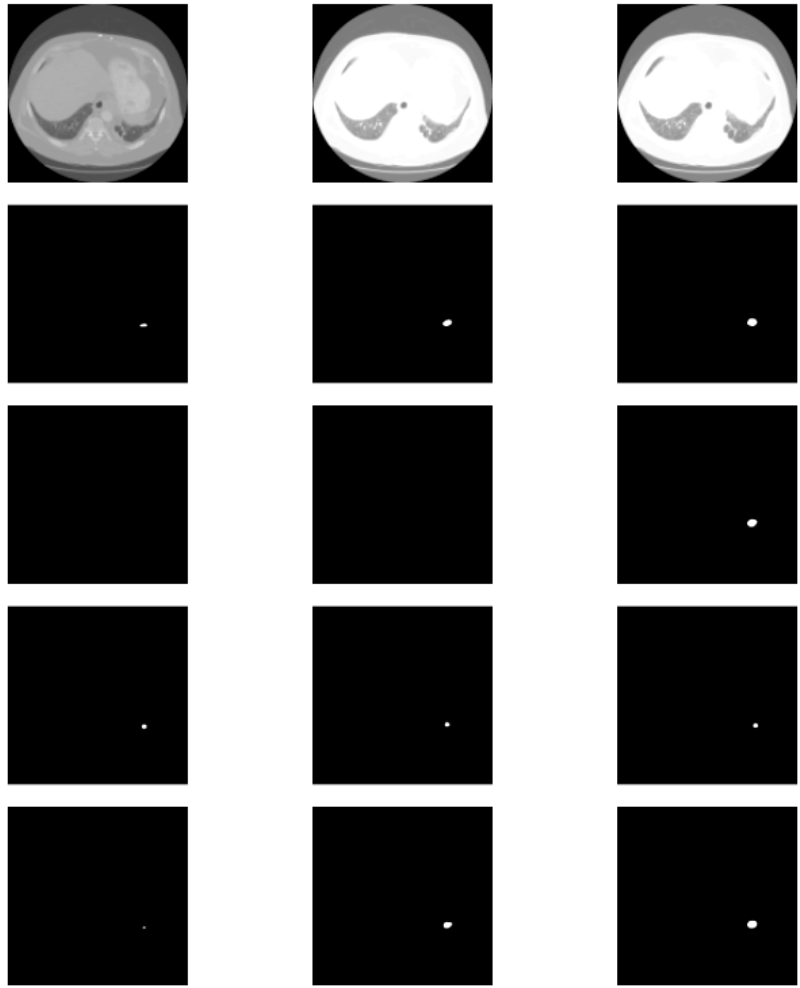

*Fig. 7.* Results of segmentation from Sytem 3 Decoder Unet++. First row — original scans. Second row — annotation. Third row — initial masks from $H$. Fourth row — positive clicks. Fifth row — corrected masks with $H \_ I$

Considering both quality and feedback criteria, we can conclude that System 3 with Unet++ decoder is a rational choice for the lung cancer segmentation task.

Finally, let's investigate the systems' performance on more interactive feedback iterations in Fig. 9.

From the received plots, System 2 outperforms System 3 in mean IoU only from the third feedback iteration while still losing in feedback score. These results are logical because System 2 always relies on expert feedback, while System 3 has initial mask approximation, which is not conditioned on feedback. On the other hand, there is still an issue with the "ideal" behavior for users' feedback simulation, and this issue will be addressed in future works.

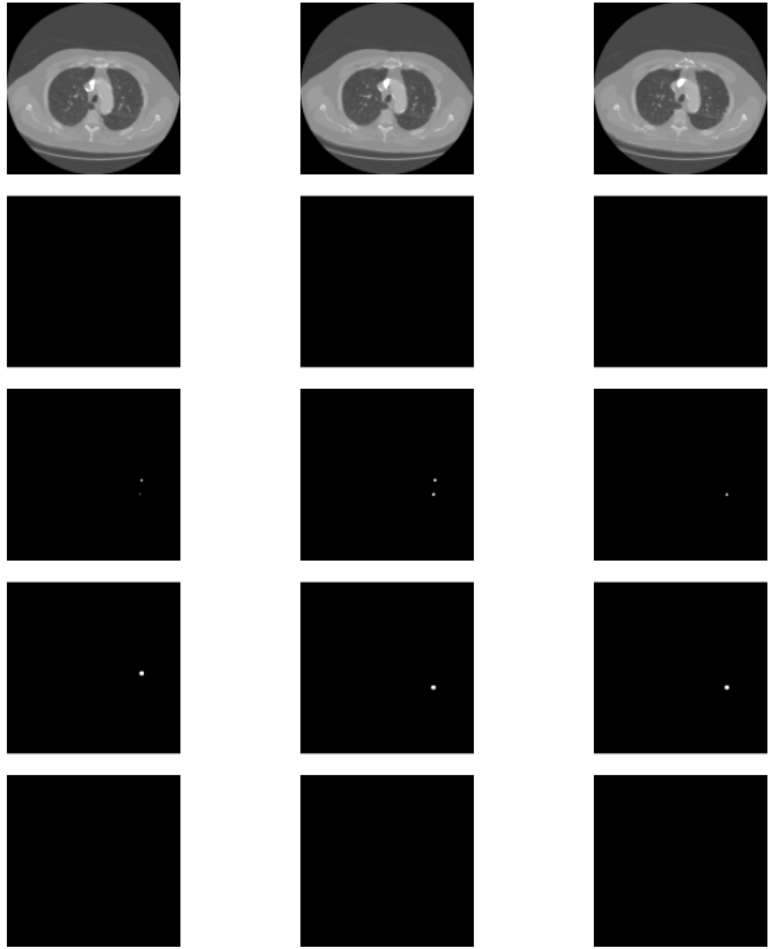

*Fig. 8.* Results of segmentation from Sytem 3 Decoder Unet++. First row — original scans. Second row — annotation. Third row — initial masks from $H$. Fourth row — negative clicks. Fifth row — corrected masks with $H\_1$

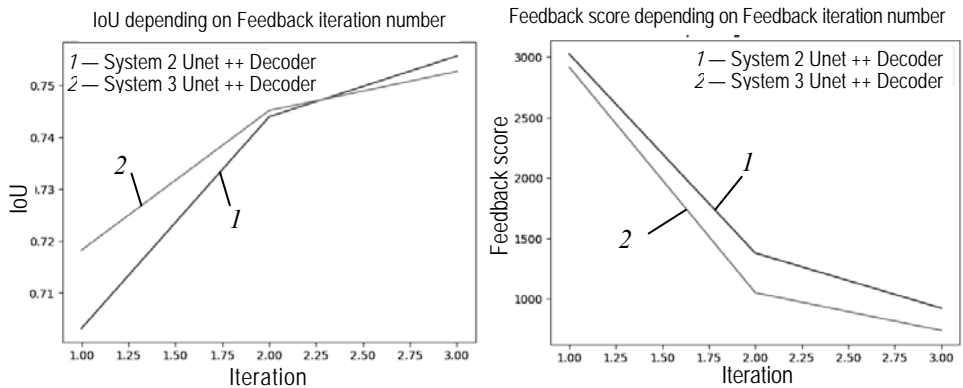

*Fig. 9.* Mean IoU and Feedback Score depending on feedback iteration number for Systems 2 and 3 with Unet++ Decoder

## CONCLUSIONS

This paper has studied different IDSS for lung cancer segmentation, proposed evaluation criteria, and the algorithm for users' feedback simulation. Finally, a system that outperforms previous approaches by all criteria is proposed. However, it may have less quality increase with feedback iterations compared to previous systems. This issue will be addressed in future works. It is important to mention that the proposed system is a combination of 2 other systems, so it is possible that there is a huge room for improvement. Another issue that requires further research is the users' feedback simulation algorithm. Its behavior is too "ideal," and there should be randomness conditioned on previous mask approximation results.

**INFORMATION ON THE ARTICLE**

**Volodymyr S. Sydorskyi,** ORCID: 0000-0001-9697-7403, National Technical University of Ukraine "Igor Sikorsky Kyiv Polytechnic Institute", Ukraine, e-mail: volodymyr.sydorskyi@gmail.com

**ІНТЕРАКТИВНА СИСТЕМА ПІДТРИМАННЯ ПРИЙНЯТТЯ РІШЕНЬ ДЛЯ СЕГМЕНТАЦІЇ РАКУ ЛЕГЕНІВ** / В.С. Сидорський

**Анотація.** Досліджено клінічні інтелектуальні системи підтримання прийняття рішень (ІСППР) для сегментації раку легень, які базуються на глибинних нейронних мережах. Запропоновано нову інтерактивну ІСППР і порівняно її з попередніми підходами. Обговорено проблему невизначеності цілей під час створення інтерактивних систем і запропоновано критерії для оцінювання якості та кількості зворотного зв'язку від експерта. Для автоматизації оцінювання системи використано спеціальний алгоритм для симуляції зворотного зв'язку експерта. Запропонована інтерактивна ІСППР перевершила попередні підходи (як інтерактивні, так і неінтерактивні) у завданні сегментації раку легень. Цей підхід перспективний як щодо якості, так і зручності використання експертом. Водночас обговорено низку можливих модифікацій, які можна виконати для покращення як критеріїв оцінювання, так і запропонованої ІСППР у майбутніх працях.

**Ключові слова:** клінічні системи підтримання прийняття рішень, глибинне навчання, відкрита система, інтерактивна сегментація.